# A Deep Belief Network Based Machine Learning System for Risky Host Detection

Wangyan Feng
Center for Advanced Machine Learning
Symantec Corporation
Mountain View, California, USA
Charles_Feng@Symantec.com

Shuning Wu
Center for Advanced Machine Learning
Symantec Corporation
Mountain View, California, USA
Shuning_Wu@Symantec.com

Xiaodan Li
Department of Electronic and Computer Engineering
Duke University
Durham, North Carolina, USA
xiaodan.li@duke.edu

Kevin Kunkle
Department of Computer Science
Indiana University
Bloomington, Indiana, USA
kevinjameskunkle@gmail.com

*Abstract*—To assure enterprise security, typically a SIEM (Security Information and Event Management) system is built to correlate security events from different preventive technologies and flag alerts. Analysts in a security operations center (SOC) investigate the alerts to decide whether the related hosts are malicious or not. However, the number of alerts is overwhelming which exceeds the SOC's capacity to handle and the false positive rate is also really high. Consequently, there is a great need to reduce the false alarms as much as possible. Instead of detecting network intrusion from outside of the enterprise, this paper focuses on detecting compromised hosts within enterprise by an intelligent Deep learning system. Our system leverages alert information, various security logs and analysts' investigation results in a real enterprise environment to identify hosts with high likelihood of being compromised. Text mining and graph-based method are used to generate targets and extract features. In order to validate the effectiveness of our model, other machine learning algorithms such as Multi-layer Neural Network, Deep Neural Network, Random Forest *etc.* are applied to the same enterprise data. The results indicate that the Deep Belief Network (DBN) performs much better than other algorithms and is 6 times more effective than the current rule-based system. What is more, due to its effectiveness, this compromised host detection system has been implemented in a real enterprise production environment, which includes data collection, label creation, feature engineering and host score generation.

## I. INTRODUCTION

Cyber security incidents, especially data breaches, will cause significant financial and reputation impacts on an enterprise. In 2015, IBM and Ponemon Institute conducted research on the cost due to data breach in 62 companies. The average cost of data breach is $6.5 million [1]. In order to detect malicious activities, a SIEM (Security Information and Event Management) system is built in companies or government. The system normalizes and correlates events logs (the raw data is organized to reduce and eliminate redundancy/error/noise) from endpoint, firewalls, intrusion detection systems, DNS (Domain Name System), DHCP (Dynamic Host Configuration Protocol), Windows event logs, VPN logs etc. A SOC (Security Operations Center) team develops use cases with a pre-determined severity based on the analyst's experiences. They are typically rule based involving one or two indicators. These rules can be network/host based or time/frequency based. Some examples are:

- Detection of multiple malware infections that cannot be cleaned by endpoint protection software

- Failed login attempts towards the same PCI (Payment Card Industry) asset exceeding a certain number

- Certain number of denied firewall events from PCI servers within a pre-specified time window

If any event triggers one or multiple use cases, the SIEM will generate an alert immediately. Then the analyst in the SOC team will investigate the alert to decide whether the host related to the alert is risky (true positive) or not (false positive). However, SIEM generates a lot of the alerts, but with a very high false positive rate. The number of alerts per day is much more than the capacity of SOC. Therefore, SOC may choose to only investigate the alerts with high severity or suppress the same type of alerts (for example, if the same type of alerts keep triggering within 7 days, then the same alerts will be ignored by SOC). This could potentially miss some severe attacks. Consequently, a more intelligent and automatic system is required to identify risky hosts.

Machine learning has been adopted in network intrusion detections and malware classifications. However, to the best of our knowledge, little research focuses on detecting compromised hosts by machine learning models. Compared to network intrusion detection, risky host detection is much more challenging. As to network intrusion detection, the features are more objective. For example, if a file creates auto-run tasks or injects code into other processes, or if denial of service attack happens on a web server, the labels are easier to create without manual inspection, i.e., endpoint protection software or network intrusion detection system can automatically catch the malicious file or intrusion behaviors. However, in risky host detection, it is related to host/user behavior, so the problem is more subjective and complicated. Usually, SOC analyst needs to spend a lot of time investigating multiple alerts from the host to determine if it is risky, hence the label generation process is highly manual and labor intensive. Our research aims at this challenge and tries to provide better decision support for the SOC analyst.

Specifically, our approach utilizes Deep Belief Network to evaluate host risk according to events logs from endpoint, firewalls, intrusion detection systems, DNS (Domain Name System), DHCP (Dynamic Host Configuration Protocol), Windows event logs, VPN logs etc. Based on this information, we will evaluate host state (risky or not). This approach can provide the security analyst a risk score for every host and the security analyst can focus on those with high risk scores.

There are two major advantages of Deep Belief Network over other machine learning models:

- Hosts' behaviors or features are very complicated. For example, a host may visit different malicious websites, download/upload different files or have different malware infections at the same time. In fact, we create hundreds of features per day for each host to describe its security posture. The greatest advantage of Deep Belief Network is its capability of "learning complex features", which is achieved by its layer-by-layer learning strategy where the higher-level features are learned from the previous layers (for example, in face recognition, higher layer learns from pixels to lines to noses to faces) hence the higher-level features better extract the information from the input data's structures.

- Hosts' labels are very limited. Traditional supervised machine learning methods rely on sufficient amount of labels to make accurate predictions, while in our problem, only few hosts (~1%) have labels, which brings a big challenge to traditional supervised learning methods. However, Deep Belief Network is a great fit for this challenge as it integrates unsupervised learning into its network training. Deep Belief Network can be viewed as a composition of unsupervised Restricted Boltzmann Machines, where each Restricted Boltzmann Machine's hidden layer serves as the visible layer for the next. This leads to a layer-by-layer unsupervised training procedure and supervised learning is only applied at the end to fine-tune the network parameters and convert the learned representation into probability predictions (for example, softmax function is used in the last layer to convert the outputs from hidden layers to probability predictions). Therefore, Deep Belief Network has less dependence on initial labels and we expect it to perform better for our problem with limited labels.

The main contribution of this paper is as follows:

- An advanced machine learning system is proposed and evaluated by real industry data from Symantec. The system can effectively reduce the resources to analyze alerts manually while at the same time enhance enterprise security.

- A novel data engineering process is provided which integrates alert information, security logs, and SOC analysts' investigation notes to generate features and labels for machine learning models.

- To the best of our knowledge, this is the first paper which applies a Deep Belief Network to evaluate host risk. In addition, when compared with other machine learning models such as Multi-layer Neural Network, Deep Neural Network, Random Forest, Support Vector Machine and Logistic Regression, Deep Belief Network shows the best performance for highly unbalanced labelled data.

The remainder of the paper is organized as follows. In Section II, related work is summarized. Section III briefly describes Restricted Boltzmann Machine and Deep Belief Network, which is the core machine learning method in our research. Section IV introduces feature engineering and label creation. In Section V, experiment results on real industry data are discussed. Section VI concludes the whole paper.

## II. RELATED WORK

### A. Conventional machine learning model

Machine learning models have been applied to detect anomalies and intrusions [2]. In [3], nine classifiers (Bayesian Network, Logistic Regression, Random Forest etc.) were compared in malicious traffic detection. In [4], fuzzy clustering was introduced to decrease the false positive rate. K-Means clustering was used towards scalable unsupervised intrusion detection [5]. Dynamic behavior models such as Hidden Markov Model were adopted to detect intrusions based on host's user profiles built from normal usage data [6].

Li et al. presented an online Support Vector Machine (SVM) with decision tree to classify host state based on network traffic behavior [7]. Chand et al. combined nine other machine learning models with SVM to obtain better performance in intrusion detection [8]. In [9], a hybrid model combining SVM, decision tree and Naïve Bayes was proposed. In [10], enhanced SVM was used for network anomaly detection. In [11], Meng compared different machine learning models including artificial neural network, SVM and decision tree for network anomaly intrusion detection. In [12], Silva et al. used neural network and SVM to automatically detect hosts that disseminate web spam.

### B. Deep Learning Model

More complicated models such as Deep Neural Network (DNN) and Deep Belief Network (DBN) have been applied to identify malicious intrusion. Min-Ju et al. adopted deep neural network for in-vehicle network intrusion detection [13].

In [14], Deep Belief Network was applied to intrusion detection and showed better performance than SVM. In [15],

Liu et al. applied extreme learning machine to the training process of Deep Belief Network and improved the model performance in network intrusion detection. Current research is mainly based on some historical simulated or experimental data instead of real industry data, such as NSL KDD [16]. What is more, little research analyzes the security state of machines based on the alert information. To the best of our knowledge, this is the first paper which applies Deep Belief Network to evaluate host risk on real industry data.

## III. Deep Belief Network

Deep Belief Network has two main differences from Deep Neural Network:

- Network topology: Deep Neural Network is a feed-forward network with more than one hidden layers. Each hidden neuron typically uses the logistic/sigmoid activation function. In contrast, Deep Belief Network has undirected connections between hidden layers composed of stacked Restricted Boltzmann Machines.

- Network training: Deep Neural Network requires labelled data in the whole backpropagation training process to adjust its weights. In contrast, Deep Belief Network uses unsupervised pre-training by contrastive divergence and then fine-tunes the weights by backpropagation.

Deep Neural Network generally needs a large amount of balanced labeled data, but the majority of industry data lack such labels. Deep Belief Network is one kind of unsupervised probabilistic generative model and is mainly constructed by stacking Restricted Boltzmann Machines. The parameters for the stacked Restricted Boltzmann Machines are trained by contrastive divergence (CD) algorithm [17]. Since CD is unsupervised learning, labeled data are not needed in this stage. In the second stage, the pre-trained network will be adjusted by supervised learning model such as softmax/logistic regression or linear classifier with gradient descent learning process [18]. However, the parameters of Deep Belief Network are almost fixed after CD and the second stage only fine-tunes the model parameters. Therefore, fewer labeled data is needed in Deep Belief Network. We will provide more details on Deep Belief Network in the following sections.

### A. Restricted Boltzmann Machine

Restricted Boltzmann Machine (RBM) is a two-layer stochastic model. As shown in Figure 1, it includes hidden layer and visible layer. The visible layer consists of visible states $V = (v_1,...,v_m)$, while hidden layer has states $H = (h_1,...,h_n)$ which cannot be measured directly. Between two layers, the states are fully connected [19]. However, there are no connections among the states within the same layer which means that the states in the same layer are mutually independent.

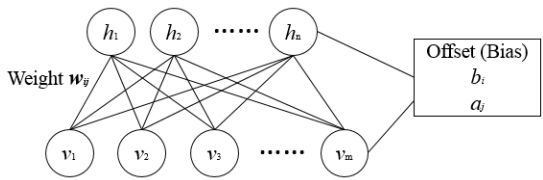

Fig. 1. Restricted Boltzmann Machine

Restricted Boltzmann Machine is one kind of energy-based model. It associates a scalar energy to each configuration of the variables of interest. The learning process involves adjusting the energy function so that its shape has desirable properties. For Restricted Boltzmann Machine, the joint probability distribution for $(v, h)$ is given by Equation (1) through an energy function:

$$p(v, h) = \frac{1}{Z} e^{-E(v,h)} \quad (1)$$

The energy function $E(v, h)$ is defined in Equation (2), where $w_{ij}$ represents the weight between hidden state $h_i$ and visible state $v_j$. $b_i$ and $a_j$ are offsets or biases as shown in Figure 1, which are constant values. The bias is used to shift energy function in the training process to get better model performance.

$$E(v, h) = -\sum_{i=1}^{n}\sum_{j=1}^{m} w_{ij} h_i v_j - \sum_{j=1}^{m} a_j v_j - \sum_{i=1}^{n} b_i h_i \quad (2)$$

The marginal probability for $v$ can be calculated by taking the sum of $e^{-E(v,h)}$ over $h$:

$$p(v) = \frac{1}{Z} \sum_h e^{-E(v,h)} \quad (3)$$

where $Z$ is the normalization factor which is the sum of all combinations of $(v, h)$:

$$Z = \sum_{v,h} e^{-E(v,h)} \quad (4)$$

In the common case of binary states (where $v_j$ and $h_i \in \{0, 1\}$), the probability that $v_j = 1$ given $h$ can be calculated from Equation (5). $\sigma(.)$ is the sigmoid function, where $\sigma(x) = 1/(1 + \exp(-x))$.

$$p(v_j = 1|h) = \sigma(a_j + \sum_i h_i w_{ij}) \quad (5)$$

Similarly, given $v$, the probability that $h_i = 1$ can be calculated from Equation (6):

$$p(h_i = 1|v) = \sigma(b_i + \sum_j v_j w_{ij}) \quad (6)$$

### B. Restricted Boltzmann Machine Training

The main task for Restricted Boltzmann Machine training is to learn the weight matrix $W = \{w_{ij}\}$ that maximizes the log likelihood $\log p(v)$. From Equations (2) and (3), we have,

$$\frac{\partial E(v, h)}{\partial w_{ij}} = -v_j h_i \implies$$

$$\frac{\partial \log p(v)}{\partial w_{ij}} = \langle v_j h_i \rangle^0 - \langle v_j h_i \rangle^\infty \implies$$

$$\Delta w_{ij} = \epsilon(\langle v_j h_i \rangle^0 - \langle v_j h_i \rangle^\infty) \quad (7)$$

where $\langle v_j h_i \rangle^t, t = 0, ..., \infty$ denotes the expectation of random variable $v_j h_i$ at sampling step $t$ and $\epsilon$ is the learning rate for weight updating. Sampling of $\langle v_j h_i \rangle^t$ can be obtained by running a Markov chain to convergence through Gibbs sampling. For Restricted Boltzmann Machine, since visible state and hidden state are conditionally independent, visible states can be sampled simultaneously given fixed values of the hidden states. Similarly, hidden states can be sampled simultaneously given the visible states. Therefore, given $h_i^t$ or $v_j^t$ at step $t$, $h_i^{t+1}$ or $v_j^{t+1}$ at step $t + 1$ can be obtained by

Equations (5) and (6). That is, $h_i^{t+1}$ is set as 1 with probability $\sigma(b_i + \sum_j v_j^t w_{ij})$, and $v_j^{t+1}$ is set at 1 with probability $\sigma(a_j + \sum_i h_i^{t+1} w_{ij})$. As $t \to \infty$, the samples will be approaching true samples from $p(v)$. Figure 2 illustrates this sampling process.

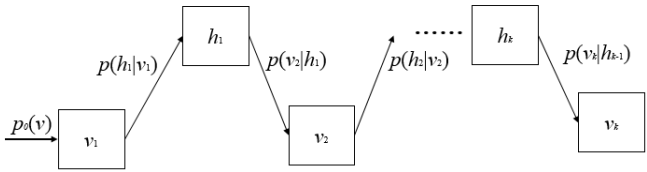

Fig. 2. Restricted Boltzmann Machine sampling process

Contrastive Divergence (CD) is an efficient way to speed up the sampling process [19]. First, instead of initiating the Markov chain randomly, it starts from training samples that are closer to the true distribution, so the chain will converge faster. Second, CD does not wait for the Markov chain to fully converge. Instead, it will stop after $k$ steps of Gibbs sampling. Previous research found that even for small $k$ (in practice, $k$ is often set as 1), the algorithm obtains close results to the final maximum likelihood solution [20].

*C. Deep Belief Network Structure*

Deep Belief Network is a probabilistic generative model. As shown in Figure 3, Deep Belief Network is primarily constructed by stacking Restricted Boltzmann Machines. It has two parts: stacked Restricted Boltzmann Machines and classifier. The training process for Deep Belief Network includes pre-training and fine-tuning respectively.

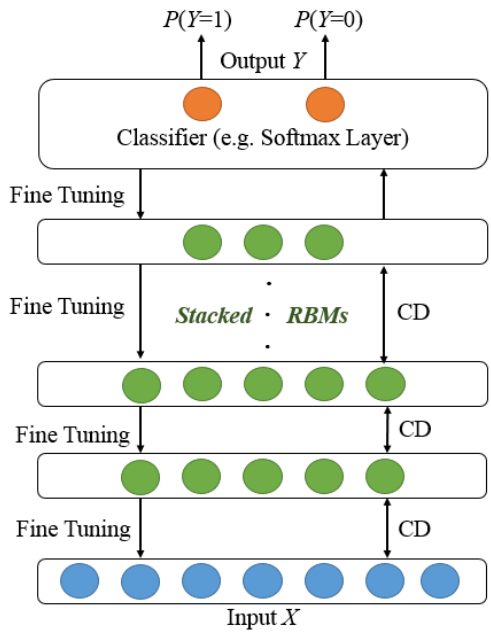

Fig. 3. Structure of Deep Belief Network

*1) Pre-training*

The pre-training is applied to stacked Restricted Boltzmann Machines. The hidden layer of the previous Restricted Boltzmann Machine will be the visible layer of the next Restricted Boltzmann Machine. Let $X$ be the input matrix, the pre-training process is described below:
1. Train first layer of Restricted Boltzmann Machine on $X$ to obtain the weight matrix using Contrastive Divergence algorithm
2. Transform $X$ with the weight matrix to reconstruct new data $X'$
3. Use $X'$ as new input, $X' \to X$, for the next two layers of Restricted Boltzmann Machines
4. Repeat Step 1 to Step 3 until the last two layers of the network are reached

*2) Fine-tuning*

In our research, backpropagation and softmax regression are used in fine-tuning of Deep Belief Network.
- Backpropagation: Backpropagation is used to adjust weights by the derivative chain principle on model errors. The error will be propagated from the last layer to the first layer. Two key parameters here are batch size and number of epochs. In fine-tuning, training samples are divided into groups with same size. Batch size is the number of samples in each group fed to the network before weight updates are performed. Number of epochs is related to the iterations of fine-tuning. Generally speaking, the network will undergo more fine-tunings with smaller batch size or larger number of epochs [21].
- Softmax regression: Softmax regression is the extension of logistic regression, which can be applied to more than two classes. The conditional probability of $Y=l$ (class $l$) given stacked Restricted Boltzmann Machines' output $X'$, coefficient matrix $C$ and intercept $d$ is shown in Equation (8). Final prediction is the class with largest probability in Equation (9).

$$P(Y = l|X', C, d) = \frac{e^{c_l^T x' + d_l}}{\sum_k e^{c_k^T x' + d_k}} \quad (8)$$

$$Prediction = max_{l \epsilon L}\{P(Y = l|X', C, d)\} \quad (9)$$

IV. FEATURE ENGINEERING AND TARGET CREATION

*A. Raw Data Description*

The raw data is collected from Symantec internal security logs. It consists of alerts from SIEM system, notes from analysts' investigation, and logs from different sources, including firewall, intrusion detection/prevention system, HTTP/FTP/DNS traffic, DHCP, vulnerability scanning, Windows security event, VPN and so on. The logs have terabytes of data each day. Table 1 lists key elements in SIEM alert data:

TABLE I. SIEM ALERT DATA ELEMENTS

| Field Name | Description |
|---|---|
| Host_ID | Host name (e.g., "Dell-PC-0001") |
| Event_ID | Security event (e.g. "SYS0206 – Malware Not Remediated") |
| Time | Event time stamp |
| Severity | Event security (e.g. 1-10) |

Analyst's investigation "notes" or annotations are usually stored in a ticketing system (such as OTRS or Resilient) as free-form text. The note contains information on whether an alert is true positive or not and we use it to create labels for machine learning dataset.

*B. Feature Engineering*

Before feeding to our model, the raw data needs to be pre-processed, since the industry data is not well-structured. Although deep learning models are able to derive features internally and automatically, we still have to generate initial features from raw data for the input layer as we normally do for conventional machine learning models. The steps of feature engineering are illustrated below:

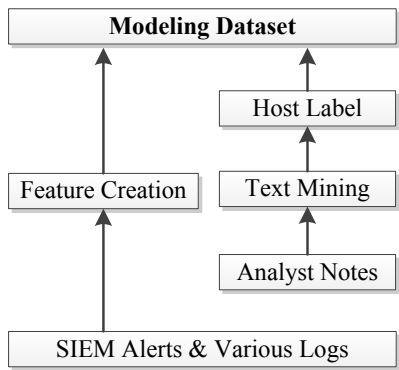

Fig. 4. Data engineering process

The features are created at the individual host level as our main goal is to predict the host's risk. We have created 100 features to describe a host. The features can be classified in the following four categories:

- Summary features: these features can be generated from statistical summaries. For example, the number of "Malware Not Remediated" event over the last 24 hours, or the number of high severity events (severity over 7) over the past 7 days.
- Indicator features: these features are in binary (0 or 1) format, for example, whether "Malware Not Remediated" event happens over weekend.
- Temporal features: these features include time information, for example, security event arrival rate, which takes into account the time interval between two consecutive events.
- Relational features: these features are derived from social graph analysis, for example, host weighted PageRank value calculated from host-event graph [22]. The nodes are hosts or events. The relationship between a host and its events is represented by edge in graph, and the edge weight is the number of certain event on a host. Below is an example of a subgraph on one certain host with its security events and some weighted PageRank values derived from the bigger graph with more host and events. Higher PageRank implies more suspicious host behaviors.

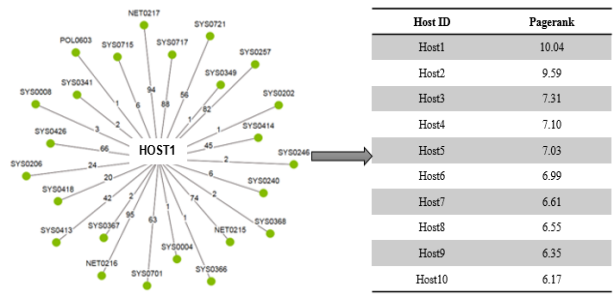

Fig. 5. Example on host weighted pagerank feature

*C. Label Generation*

After all features are generated, we need to assign targets or "labels" for our machine learning models. The labels are obtained by mining analyst's investigation notes, including but not limited to:

- Initial background: the reasons why an alert was triggered.
- Internal research: support information from different internal system logs.
- External research: support information from external resources, such as IP geolocation and reputation.
- Investigation result: whether the event is non-malicious, false positive, or escalated.

Text mining techniques, such as key word/topic extraction and sentiment analysis, are required to extract the host's actual state. Here are some examples from text mining:

- Topic 1: "Unable to gather supporting evidence for this alert" → Host status from text mining: normal
- Topic 2: "Host made connections with malicious domains" → Host status from text mining: risky
- Topic 3: "Advanced malware infection detected on this host" → Host status from text mining: risky

Finally, we attach the labels from text mining as the targets for our machine learning models. The final analytic dataset will look like this:

TABLE II. EXAMPLE ON FINAL MODELING DATASET

| Host ID | Summary feature 1 | Indicator feature 2 | Temporal feature 3 | Relational feature 4 | … | Label |
|---|---|---|---|---|---|---|
| Host1 | 13 | 1 | 0.65 | 5.17 | … | 1 (risky) |
| Host2 | 25 | 0 | 2.74 | 9.34 | … | 1 (risky) |
| Host3 | 4 | 0 | 1.33 | 3.52 | … | 0 (normal) |

## V. EXPERIMENT RESULTS

*A. Model Performance Measures*

The dataset includes one month's security events with 20,572 distinct hosts and 100 features. The dataset is highly unbalanced with only ~1% of host verified as risky by security analyst. Therefore, current rule-based system is not very effective and generates a lot of false positives. Security analysts spend a lot of time investigating non-risky hosts. The unbalanced data also significantly increases the challenges to predict risky hosts for machine learning models.

As a common practice, we split the data randomly into training (75% of the samples) and testing (remaining 25%) sets. We tried different training/test split ratios between training and test data (i.e., 50% / 50%, 60% / 40% and 75% / 25%), but we noticed that the split ratios had little impact on classification results. Due to the space limitation, we only discuss the experiment results with 75% / 25% split. In the testing data, we have 5,143 hosts but only 60 of them (1.17%) have been verified as risky by security analyst. We will test different machine learning models to see if they can perform better than current rule-based system. The measures of goodness are defined in Equations (10) to (12) below:

Model AUC
$$\approx \sum_i \frac{y_{i+1}+y_i}{2} \times (x_{i+1} - x_i),$$
where $(x_i, y_i)$ is the point on ROC curve  (10)

Model Detection Rate
$$= \frac{\text{Number of Risky Hosts in Certain Predictions}}{\text{Total Number of Risky Hosts}} \times 100\%$$ (11)

Model Lift
$$= \frac{\text{Proportion of Risky Hosts in Certain Predictions}}{\text{Overall Proportion of Risky Hosts}}$$ (12)

Area under ROC Curve (AUC) evaluates model's overall accuracy by approximating the region under the Receiver Operating Characteristic (ROC) curve. ROC curve displays the tradeoff between true positive rate (TPR) and false positive rate (FPR) of a classifier, where TPR is plotted along the $y$ axis and FPR is plotted along the $x$ axis. If the model is simply random guessing, its AUC value will be 0.5. The closer the ROC gets to the optimal point of perfect prediction (where FPR=0 and TPR=1), the closer the AUC gets to 1.

Different from AUC that evaluates model on the whole test data, detection rate and lift reflect how good the model is in discovering risky hosts among different portions of predictions. To calculate these two metrics, the results are first sorted by the model scores (the probability of a host being risky in our case) in descending order. Detection rate measures the effectiveness of a classification model as the ratio between the results obtained with and without the model. For example, suppose we take the top 10% of the predictions, and the model captures 30 actual risky hosts, the detection rate for top 10% predictions is equal to 30/60=50% (note that we have verified 60 risky hosts in the test data). If we do this for 20%, 30%, etc. and then plot the detection rates from different portions of predictions, we get a detection rate chart. Similar to ROC curve, the better the classification model is, the steeper the line will be in detection rate chart.

Lift measures how many times it is better to use a model in contrast to not using a model. Using the same example above, we have 30 risky hosts captured in top 10% predictions, the lift is equal to (30/514)/(60/5143)≈5. Assuming an unpredictive model that is no better than current rule-based system, the baseline lift value will be 1. Similar to detection rate chart, we can vary the portions of predictions and plot the lift values on a lift chart. Higher lift implies better performance from a model on certain predictions.

B. *Model Parameters*

The settings of Deep Belief Network are:

- Number of epochs for fine-tuning is 100.

- Four layers with one input layer, two hidden layers and one output layer. Each layer has 100, 20, 10, 2 neurons respectively. Note that adding more hidden neurons may lead to model over fitting so it will not be helpful to the model performance. Figure 6 shows the simulation over different number of hidden neurons $N$, where the layer neurons vary over (100, $N$, $N/2$, 2). It can be seen that model training time increases almost linearly with $N$, which implies that the model has good scalability with more hidden neurons. On the other hand, AUC on test data goes up when $N<25$ and starts to drop after that, which may indicate that the network's generalizability decreases with too many hidden neurons. We tried more complicated structure with more neurons and layers, but there is no obvious performance improvement, so we decide to go with simpler network structure.

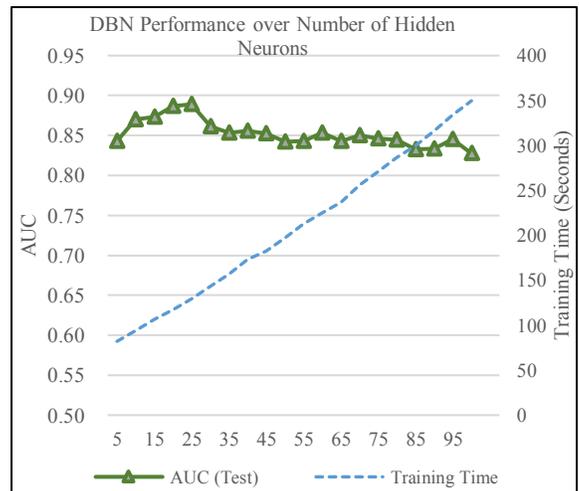
Fig. 6. AUC/training time with respect to number of hidden neurons

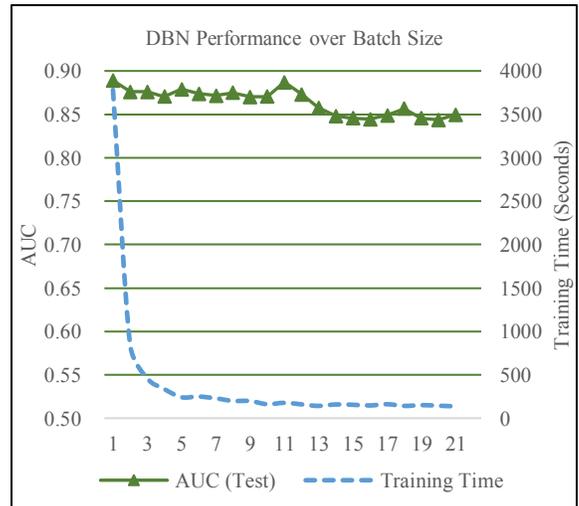
Fig. 7. AUC/training time with respect to DBN batch size

- Batch size is set as 50. For smaller batch size, the model may fit the data better but has higher computational cost. From some simulations shown in Figure 7, we find that the training time has a steeper fall when batch size is less than 50, then it gets steady after that. Also AUC on test data tends to drop with larger batch size. AUC is close to the best value when batch size is 50. Therefore, it is reasonable to set batch size as 50 with good balance between training time and model accuracy.

Other machine learning models include:

- Multi-layer Neural Network (MNN): Three layers with one input layer, one hidden layer and one output layer. Each layer has 100, 50, 2 neurons respectively.
- Deep Neural Network (DNN): Four layers with one input layer, two hidden layers and one output layer. Each layer has 100, 20, 10, 2 neurons respectively. The layer setting is the same as Deep Belief Network so as to compare the performance between these two deep learning methods.
- Random Forest (RF): 100 decision trees with entropy-based splits.
- Support Vector Machine (SVM): with radial basis function kernel.
- Logistic Regression (LR): generalized linear model with Binomial family and Logit link function.

### C. Experiment Results

The first thing we notice in our experiment is that Deep Neural Network does not work well for our problem, which has lowest AUC value on test dataset. We will discuss this further in Section VI. In this section, some abbreviations will be used: "DBN" for Deep Belief Network, "MNN" for Multi-layer Neural Network, "RF" for Random Forest, "SVM" for Support Vector Machine, "LR" for Logistic Regression, and "DNN" for Deep Neural Network.

The table below lists AUC of different models on test data. We vary the random samples in the training and test data and both mean and standard error of AUC are provided in Table III. As we expected, DBN has highest AUC among all models. Generally, AUC over 0.8 indicates that the model works fairly well. With average AUC value close to **0.85**, DBN achieves satisfying accuracy on risky host detection.

In addition, SVM and DNN have much higher standard errors on AUC, implying that these two methods are more sensitive to training samples. On the other hand, DBN has highest average AUC value with relatively small standard error. This shows that DBN is not only more accurate than other models but also more robust to the variations of input data. Figure 8 shows the ROC curves corresponding to the best AUC values from different models.

TABLE III. MODEL AUC ON TEST DATA

|  | DBN | MNN | RF | SVM | LR | DNN |
|---|---|---|---|---|---|---|
| MEAN | **0.844** | 0.807 | 0.829 | 0.775 | 0.754 | 0.624 |
| S.D. | **0.008** | 0.006 | 0.004 | 0.016 | 0.008 | 0.020 |

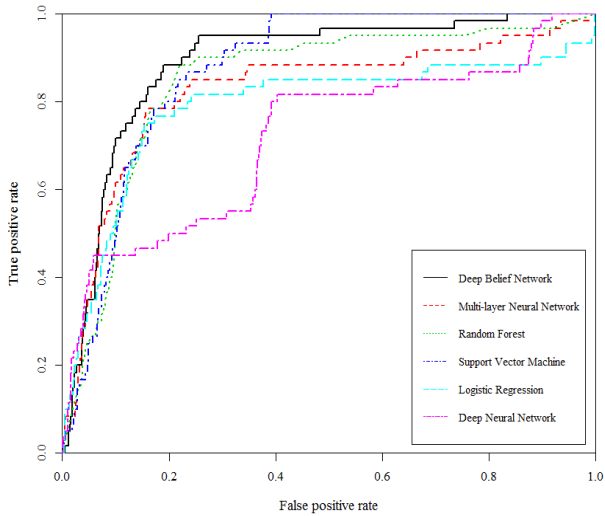

Fig. 8. ROC curves from different models (best AUC values)

As to detection rate and lift, we analyzed the best results from each model on different test data. Table IV lists the detection rates for different models on top 5% to 20% predictions respectively. Considering that we only have 1.17% risky hosts in this test dataset, it is promising that DBN is able to detect over **88%** of the true risky cases with only 20% highest predictions - more than 8 points higher than other models. Again we can find out that DNN performs worse than other models.

TABLE IV. MODEL DETECTION RATES ON TOP 5%~20% PREDICTIONS

| Top % Predictions | DBN | MNN | RF | SVM | LR | DNN |
|---|---|---|---|---|---|---|
| 5% | **35.00%** | 31.67% | 25.00% | 20.00% | 31.67% | 38.33% |
| 10% | **65.00%** | 58.33% | 43.33% | 46.67% | 50.00% | 45.00% |
| 15% | **78.33%** | 70.00% | 70.00% | 70.00% | 68.33% | 46.67% |
| 20% | **88.33%** | 78.33% | 80.00% | 80.00% | 76.67% | 48.33% |

The detection rates with respect to different prediction proportions are shown in Figure 9. DBN is much better than other models until the prediction proportion reaches top 40%. Since the predictions will be used as host risk ranking scores, DBN is more valuable as it detects more risky hosts with less predictions. This is very helpful as SOC analyst has limited resources for investigation.

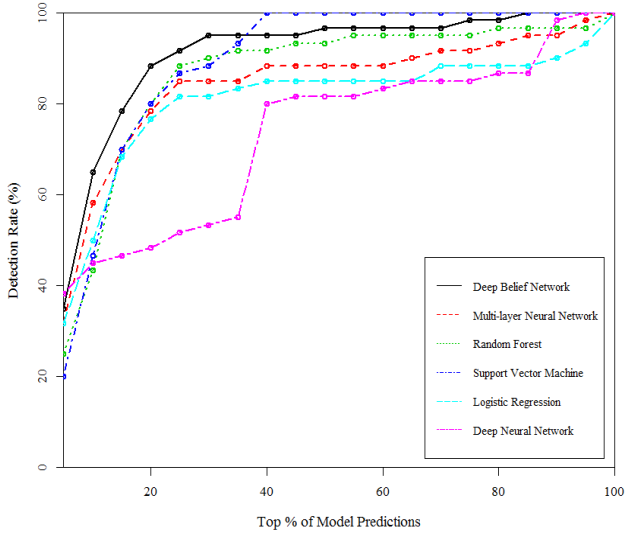

Fig. 9. Model detection rates on different proportions of predictions

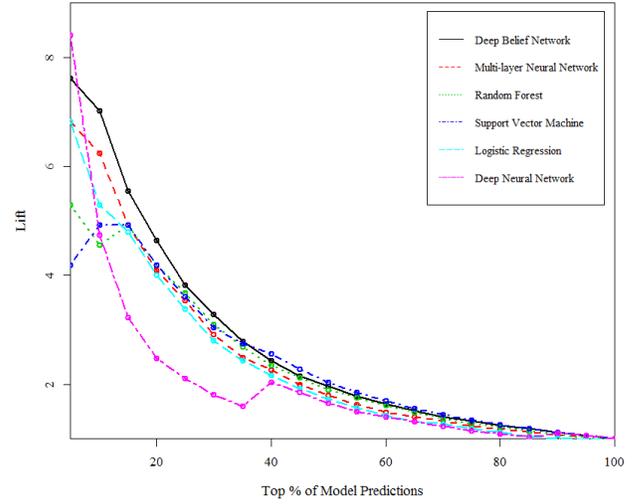

Fig. 10. Model lifts on different proportions of predictions

Finally, we evaluate model lift with top 5% to 20% predictions. The result is listed in Table V. For top 5% predictions, the lift value of DBN is 7.61, meaning that its performance is **7 times** better than current rule-based system. DNN also gets good lift value as 8.40 for top 5% predictions, but its performance degrades quickly. For example, with top 15% predictions, DNN's lift value is worse than other five models, while DBN still performs really well.

Moreover, DBN tends to capture more risky hosts in higher predictions, which can be seen from the monotonically descending lift values. RF and SVM, on the other hand, may place some risky hosts in lower prediction buckets. For example, the lift value of RF at top 15% predictions (4.92) is higher than the lift value at top 10% predictions (4.55), so the percentage of risky hosts in top 10% predictions is actually lower than that in top 15% predictions, indicating that some risky hosts cannot be captured in top 10% predictions and the model needs to include more predictions to cover these risky hosts, which is less useful in practice since SOC analyst is only able to look at less than 10% suspicious hosts.

If we look at the average lifts on top 5% to 20% predictions, DBN is the highest which is over 6 as listed on the last row of Table V. This is very promising. The model lifts on different proportions of predictions are shown in Figure 10. Generally speaking, DBN performs the best, followed by MNN and LR.

TABLE V. MODEL LIFTS ON TOP 5%~20% PREDICTIONS

| Top % of Predictions | DBN | MNN | RF | SVM | LR | DNN |
|---|---|---|---|---|---|---|
| 5% | **7.61** | 6.82 | 5.30 | 4.19 | 6.82 | 8.40 |
| 10% | **7.01** | 6.25 | 4.55 | 4.92 | 5.30 | 4.74 |
| 15% | **5.55** | 4.92 | 4.92 | 4.92 | 4.80 | 3.22 |
| 20% | **4.65** | 4.09 | 4.19 | 4.19 | 4.00 | 2.48 |
| Average | *6.21* | *5.52* | *4.74* | *4.56* | *5.23* | *4.71* |

### D. Model Implementations

Currently the proposed machine learning system has been implemented in a real enterprise production environment as illustrated in Figure 11.

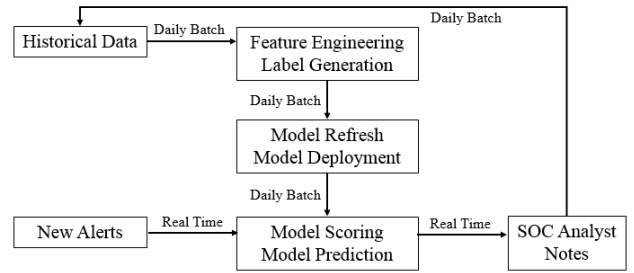

Fig. 11. Model implements in production system

The features and labels are being updated daily from historical data. Then the machine learning model is refreshed and deployed to the scoring engine daily to make sure it captures the latest patterns from the data. After that, the risk scores are generated in real time when new alerts are triggered, so SOC analysts can take action right away for high risk hosts. Finally, SOC analysts' notes will be collected and fed back into historical data for future model refinement. The whole process has been streamlined automatically from data integration to score generation. The system also actively learns new insights generated from analysts' investigations.

Regarding computational cost, we run the machine learning system on a comodity workstation (4-core Xeon CPU with 32GB ram). The whole Deep Belief Network training and scoring process can be completed in less than 30 minutes. As the machine learning system is running on a daily basis, the computational cost is trivial.

## VI. Conclusions and Discussions

Different from previous research on network intrusion detection, a novel approach based on Deep Belief Network is proposed, which leverages big data of various security logs, alert information, and analyst insights to the identification of risky host. In order to validate the effectiveness of our model, other classifiers are also applied to the same dataset. The result shows that Deep Belief Network is able to learn more insights from the highly unbalanced data than traditional machine learning models. Its average AUC on test data is close to 0.85 and average lift on top 20% predictions is over 6 times better than current rule-based system. The whole machine learning system is implemented in production environment to score and flag risky hosts in real time while updating itself by incorporating analysts' feedbacks from investigations. The system is fully automated from data acquisition, daily model refreshing, to real time scoring, which greatly improve SOC analyst's efficiency and enhance enterprise risk detection and management.

From the experiments, we notice that Deep Neural Network does not perform very well on this problem, while it has been very successful in other fields such as image recognition applications. The reasons are:

- Image recognition has a large amount of balanced labelled samples, while in our problem the dataset is smaller and highly unbalanced. Deep Neural Network can easily get stuck in a local optimum, leading to high generalization of errors and over-fitting as the network structure becomes more complicated.

- Deep Belief Network, on the other hand, does not rely on labels as much as Deep Neural Network. It mainly takes advantage of unsupervised learning in the pre-training process to train the network structure, and labels are only used later to fine-tune the structure parameters with back-propagation. Hence Deep Belief Network is not only capable of learning non-linear relationships by deep network structure, but also more robust to overfitting than Deep Neural Network.

As to the future work, we will incorporate more information from enterprise network and add more features to improve the detection accuracy and at the same time decrease false positive rate. Deep Belief Network will also be combined with other deep learning models such as Convolutional Neural Network (CNN) to achieve better classification performance.

## Acknowledgment

We would like to express our gratitude to Global Security Office of Symantec for providing data, feedback, collaboration and support to our research. Their insight and expertise greatly assisted the research. We also thank our colleague, Ningwei Liu, for his help on data engineering and model refinement.